\documentclass[12pt]{article}
\usepackage{amssymb}
\newcommand{\vek}[1]{\mbox{\bf #1}}
\begin{document}
\section*{}
\hspace*{\fill} LMU--TPW--97/02\\
\hspace*{\fill} hep--th/9701047 \\[3ex]

\begin{center}
\Large\bf
The Coulomb potential in gauge theory with a dilaton
\end{center}
\vspace{2ex}
\normalsize \rm
\begin{center}
{\bf Rainer Dick}\\[0.5ex] {\small\it
Sektion Physik der Universit\"at M\"unchen\\
Theresienstr.\ 37, 80333 M\"unchen,
Germany}
\vspace{3ex}
\end{center}

\noindent
{\bf Abstract:} I calculate the potential of a pointlike particle
carrying  SU$(N_c)$ charge in a gauge theory with a dilaton. 
The solution depends on boundary conditions imposed on
the dilaton: For a 
dilaton that vanishes at infinity the resulting potential
is of the form $(r+r_\phi)^{-1}$, with $r_\phi$ inverse proportional to the
decay constant of the dilaton.
Another natural constraint for the dilaton $\phi$ is independence
of $\frac{1}{g^2}\exp(\frac{\phi}{f_\phi})$ from the gauge coupling $g$.
This requirement yields a potential proportional to $r$ and 
makes it impossible
to create an isolated SU$(N_c)$ charge. 

\newpage  
\noindent
{\bf 1. }Dilatons are scalar particles
predicted by string theory
and any theory involving a compactification scale \cite{GSW}.
An unambiguous property of a string or Kaluza--Klein dilaton $\phi$ 
in four dimensions besides its scalar transformation behaviour
is its coupling to gauge fields through a term $\exp(\phi/f_{\phi})F^2$,
and I will use this as the defining property of a dilaton.
Dilatons arise from the massless spectrum of fundamental strings
in two different ways: On the one hand there is the model independent
dilaton arising as a unique massless scalar state of closed superstrings,
while on the other hand non--linear combinations of ten--dimensional
massless tensor
states measure the volume of internal dimensions and appear as 
Kaluza--Klein dilatons
in four dimensions. Linear combinations of these dilatons may couple in 
different
ways to gauge fields arising from different sectors of string theory 
and from
compactifications\footnote{The Kaluza--Klein gauge fields 
disappear from the 
low energy theory
if the compactification has a direct product structure
globally.} and the dilaton sector of 
low energy quantum field theories inherited from string theory
can become quite complicated. In spite of these possible complications
my main interest in the present paper concerns the discussion 
of a single low energy
dilaton coupling to SU($N_c$) gauge fields, in order to acquire
a better understanding
of the impact of dilatonic degrees of freedom in gauge theory.

Besides its appearance in the spectrum of string theory and Kaluza--Klein
theories, the present investigation was also motivated
by the fact that the dilaton begins to play an even more prominent r\^{o}le
through its covariance under duality symmetries:
Axion--dilaton--photon systems exhibit
a non--linear duality symmetry mixing axions and dilatons
through SL$(2,\mathbb{R})$ transformations \cite{STW} and
recent developments in string theory indicate that 
S--duality symmetries should be a generic feature of the
kinetic sector of low energy
quantum field theories \cite{SD}.

It is familiar from the axion and apparent for the dilaton 
that a (pseudo--)scalar can be very light,
yet very hard to observe if its decay constant is very large.
A relevant problem then
is the question how
a light dilaton affects the Coulomb potential and its non--abelian analog.
In order to study this problem 
we will neglect any dilaton mass in the sequel and study the
dilaton--gluon field generated by a pointlike quark. For a dilaton 
vanishing at infinity 
we will find a modified Coulomb potential
which is regularized at a radius 
\[
r_\phi=\frac{g}{8\pi f_\phi}\sqrt{\frac{1}{2}-\frac{1}{2N_c}}
\]
for gauge group SU$(N_c)$ and
\[
r_\phi=\frac{g}{8\pi f_\phi}
\]
for U(1). 
On the other hand, a non--vanishing expectation value of the dilaton
rescales the gauge coupling, giving rise to the requirement
that $ \exp(\frac{\phi}{f_{\phi}})$ should scale like $g^2$.
The unique solution satisfying this requirement yields a potential
proportional to the distance $r$ from the source.

Throughout this paper I will use the language of QCD for gauge fields, 
charges
and fermions. The translation of the results
to an abelian gauge group is straightforward.
I use 
letters from the middle of the alphabet both for spatial Minkowski
space indices and colour indices, while letters from the beginning
of the alphabet denote Lie algebra indices.\\[2ex]
{\bf 2. }We are interested in SU($N_c$) gauge theory 
augmented with a dilaton, and the theory is
described by a Lagrange density
\begin{equation}\label{lagden}
{\cal L}=-\frac{1}{4}\exp(\frac{\phi}{f_\phi})F_{\mu\nu}{}^a F^{\mu\nu}{}_a
-\frac{1}{2}\partial^\mu\phi\cdot\partial_\mu\phi
+\sum_{f=1}^{N_f}\overline{q}_f
(i\gamma^\mu\partial_\mu +g\gamma^\mu A_\mu{}^a X_a -m_f)q_f,
\end{equation}
with $X_a$ denoting a defining $N_c$--dimensional representation
of su$(N_c)$. 

This theory would arise e.g.\ through spontaneous 
compactification of five--dimen\-sio\-nal
QCD with $N_f$ quark flavours, see \cite{rd1}, with one exception: 
I did not include
couplings of the dilaton to quark masses, since in the standard
model masses are generated 
at the weak scale, far below any string or compactification scale.
I also set the axion already to zero, since the static pointlike source
considered below does not excite the axion field.

The equations of motion are
\[
\partial_\mu\Big(\exp(\frac{\phi}{f_\phi})F^{\mu\nu}{}_a\Big)
+g\exp(\frac{\phi}{f_\phi})A_\mu{}^b f_{ab}{}^c F^{\mu\nu}{}_c=
-g\overline{q}\gamma^\nu X_aq,
\]
\[
\partial^2\phi=
\frac{1}{4f_\phi}\exp(\frac{\phi}{f_\phi})F_{\mu\nu}{}^a F^{\mu\nu}{}_a,
\]
\[
(i\gamma^\mu\partial_\mu +g\gamma^\mu A_\mu{}^a X_a-m)q=0,
\]
where here and in the sequel flavour indices are suppressed.
To discuss the impact of the dilaton on the Coulomb potential 
we consider the Gauss law
and Faraday's law for stationary configurations
in the presence of a dilaton:
\[
\nabla\cdot\Big(\exp(\frac{\phi}{f_\phi})\vek{E}\Big)
-ig\exp(\frac{\phi}{f_\phi})(\vek{A}\cdot\vek{E}
-\vek{E}\cdot\vek{A})=\varrho,
\]
\[
\nabla\times\vek{E}-ig(\vek{A}\times\vek{E}
+\vek{E}\times\vek{A})=0,
\]
where in the gauge theory 
above $\varrho=g(q^+\cdot X_a\cdot q)X^a$
and $E_j=-F_{0j}{}^a X_a$.

The vector potential $\vek{A}$ is pure gauge
in the static limit
since the chromo--magnetic field
\[
\vek{B}=\nabla\times\vek{A} -ig\vek{A}\times\vek{A}
\]
vanishes, and we end up with
\begin{equation}\label{gl}
\nabla\cdot\Big(\exp(\frac{\phi}{f_\phi})\vek{E}\Big)=\varrho,
\end{equation}
\begin{equation}\label{fl}
\nabla\times\vek{E}=0.
\end{equation}

Our aim is to determine the chromo--electric potential 
for a point charge
\[
\varrho_a(\vek{r})=gC_a\delta(\vek{r})
\]
where $C_a$ denotes the expectation value of the generator $X_a$
in colour space. 
 From the relation
\begin{equation}\label{su3}
(X_a)_{ij}(X^a)_{kl}=
\frac{1}{2}\delta_{il}\delta_{jk}-\frac{1}{2N_c}\delta_{ij}\delta_{kl}
\end{equation}
one finds for arbitrary colour content
\[
\sum_{a=1}^{N_c^2-1} C_a^2=\frac{N_c-1}{2N_c}.
\]

We thus want to determine the field of a stationary pointlike quark from
\begin{equation}\label{gl2}
\nabla\cdot\Big(\exp(\frac{\phi(\vek{r})}{f_\phi})\vek{E}_a(\vek{r})\Big)
=gC_a\delta(\vek{r}),
\end{equation}
\begin{equation}\label{fl2}
\nabla\times\vek{E}_a(\vek{r})=0,
\end{equation}
and
\begin{equation}\label{dil}
\Delta\phi(\vek{r})=-\frac{1}{2f_\phi}
\exp(\frac{\phi(\vek{r})}{f_\phi})\vek{E}_a(\vek{r})
\cdot\vek{E}^a(\vek{r}).
\end{equation}

The unique radially symmetric solution to (\ref{gl2}) can be written down 
immediately:
\begin{equation}\label{sol1}
\exp(\frac{\phi(r)}{f_\phi})\vek{E}_a(\vek{r})
=\exp(\frac{\phi(r)}{f_\phi})E_a(r)\vek{e}_r=
\frac{gC_a}{4\pi r^2}\vek{e}_r
\end{equation}
whence equation (\ref{fl2}) is also satisfied. Equation (\ref{dil}) then 
translates into
\begin{equation}\label{dil2}
\frac{d^2}{dr^2}\phi(r)+\frac{2}{r}\frac{d}{dr}\phi(r)=
-\frac{g^2}{64\pi^2f_\phi}\Big(1-\frac{1}{N_c}\Big)
\exp\!\Big(-\frac{\phi(r)}{f_\phi}\Big)\frac{1}{r^4}.
\end{equation}

The form of this equation suggests an ansatz 
$\frac{\phi(r)}{f_\phi}=a\ln(\frac{r}{b})$,
which yields the solution discussed in the next section.
However, we can solve (\ref{dil2}) for arbitrary boundary conditions 
through a substitution 
\begin{equation}\label{sub}
\xi=\frac{g}{4\pi f_\phi r}\sqrt{\frac{1}{2}-\frac{1}{2N_c}}, \qquad
\psi(\xi)=\frac{\phi(r)}{f_\phi},
\end{equation}
yielding
\begin{equation}\label{eqpsi}
\frac{d^2}{d\xi^2}\psi(\xi)=-\frac{1}{2}\exp(-\psi(\xi)),
\end{equation}
or in terms of boundary conditions at infinity:
\begin{equation}\label{eqpsi2}
\psi'(\xi)^2-\psi'(0)^2=\exp(-\psi(\xi))-\exp(-\psi(0)),
\end{equation}
\[
\xi=
\int_{\psi(0)}^{\psi(\xi)}
\frac{d\psi}{\sqrt{\exp(-\psi)-\exp(-\psi(0))+\psi'(0)^2}},
\]
where a sign ambiguity has been resolved by the requirement that the dilaton
should not diverge at finite radius. The integral can be done elementary, 
with
two branches depending on the sign of $\psi'(0)^2-\exp(-\psi(0))$.

The presence of the dilaton introduced a two--fold ambiguity in the
Coulomb problem, and we have to determine from physical requirements
which boundary conditions to chose.

Here we require
that the dilaton generated by the pointlike quark vanishes at infinity 
while the
gradient satisfies the minimality condition
\begin{equation}\label{mincon}
\lim_{r\to \infty}r^2\frac{d}{dr}\phi(r)=
-\frac{g}{4\pi}\sqrt{\frac{1}{2}-\frac{1}{2N_c}}.
\end{equation}
This gives minimal kinetic energy for the dilaton at infinity subject to the 
constraint
that the 
chromo--electric field does not develop a singularity for positive finite $r$. 
Then we find for the radial dependence of the dilaton and the electric field
\begin{equation}\label{dilsol}
\phi(r)=2f_\phi
\ln\!\Big(1+\frac{g}{8\pi f_\phi r}\sqrt{\frac{1}{2}-\frac{1}{2N_c}}\Big),
\end{equation}
\begin{equation}\label{esol}
\vek{E}_a(\vek{r})=
\frac{gC_a}{4\pi\Big(r+\frac{g}{8\pi f_\phi}
\sqrt{\frac{1}{2}-\frac{1}{2N_c}}\Big)^2}\vek{e}_r,
\end{equation}
implying a modified Coulomb potential
\begin{equation}\label{coulpot}
\Phi_a(r)=
\frac{gC_a}{4\pi r+\frac{g}{2 f_\phi}\sqrt{\frac{1}{2}-\frac{1}{2N_c}}}.
\end{equation}

The result for gauge group U(1) is received through the 
substitution $N_c\to -1$.

This removal of the short distance singularity in the 
chromo--electric field would imply
finite energy of the dilaton--gluon configuration:
\begin{equation}\label{selferg}
E=\int d^3\vek{r}
\Big(\frac{1}{2}\nabla\phi\cdot\nabla\phi
+\frac{1}{2}\exp(\frac{\phi(\vek{r})}{f_\phi})
\vek{E}_a(\vek{r})\cdot\vek{E}^a(\vek{r})\Big)
=2gf_\phi\sqrt{\frac{1}{2}-\frac{1}{2N_c}}.
\end{equation}

\noindent
{\bf 3. }However, there exists another quite intriguing solution if 
we require
that $\frac{1}{g^2}\exp(\frac{\phi}{f_\phi})$ is independent of $g$.
This requirement arises naturally in string theory, since the 
non--perturbatively
fixed expectation value of the dilaton itself
is supposed to determine the coupling. In the
action (\ref{lagden}) this requirement amounts to the constraint that
the solution should respect the scale invariance of the equations
of motion under
\[
\phi\to\phi+2\zeta f_\phi
\]
\[
A\to \exp(-\zeta)A
\]
\[
g\to \exp(\zeta)g
\] 
for constant $\zeta$.
Eqs.\ (\ref{sub},\ref{eqpsi2}) then imply 
$\psi'(\xi)^2=\exp(-\psi(\xi))=4\xi^{-2}$, 
yielding
\begin{equation}\label{dilsol2}
\phi(r)=
2f_\phi\ln\!\Big(\frac{g}{8\pi f_\phi r}\sqrt{\frac{1}{2}-\frac{1}{2N_c}}\Big),
\end{equation}
\begin{equation}\label{esol2}
\vek{E}_a(\vek{r})=
\frac{32\pi f_\phi^2}{g}\frac{N_c}{N_c-1}C_a\vek{e}_r.
\end{equation}
This corresponds to an energy density 
\[
{\cal H}(\vek{r})=4\frac{f_\phi^2}{r^2}
\]
whence the energy in a volume of radius $r$ diverges linearly:
\[
E|_r=16\pi f_\phi^2 r.
\]
This is an infrared divergence which we cannot expect to be related
to new physics at short distances, and
it would cost an infinite amount of energy to create an isolated quark.\\[1ex]
{\bf Note added:} The abelian version of  the solution 
(\ref{dilsol})--(\ref{coulpot})
was introduced as a solitonic excitation of the dilaton--photon system
in a remarkable paper
by Cveti\v{c} and Tseytlin \cite{CT}.
I would like to thank A.\ Tseytlin for drawing my attention to this work.


\begin{thebibliography}{88}
 \bibitem[1]{GSW} M.B.\ Green, J.H.\ Schwarz and 
  E.\ Witten, {\it Superstring Theory},
  2 Vols., Cambridge University Press, Cambridge 1987.
 \bibitem[2]{STW} A.\ Shapere, S.\ Trivedi and F.\ Wilczek,  
 {\sl Mod.\ Phys.\ Lett.\ }{\bf A6} (1991) 2677.
 \bibitem[3]{SD} A.\ Font, L.\ Iba\~{n}ez, D.\ L\"ust and F.\ Quevedo,
 {\sl Phys.\ Lett.\ }{\bf B249} (1990) 35;\\
 J.H.\ Schwarz and A.\ Sen, {\sl Nucl.\ Phys.\ }{\bf B411} (1994) 35,
 {\sl Phys.\ Lett.\ }{\bf B312} (1993) 105;\\
 A.\ Sen, {\sl Int.\ J.\ Mod.\ Phys.\ }{\bf A9} (1994) 3707,
 {\sl Phys.\ Lett.\ }{\bf B329} (1994) 217;\\
 J.H.\ Schwarz, {\it Lectures on superstring and M theory dualities},
 hep--th/9607201.
 \bibitem[4]{rd1} R.\ Dick, {\sl Phys.\ Lett.\ }{\bf B380} (1996) 29. 
 \bibitem[5]{CT} M.\ Cveti\v{c} and A.A.\ Tseytlin, 
 {\sl Nucl.\ Phys.\ }{\bf B416}
 (1994) 137.

\end{thebibliography}
\end{document}